\documentclass[useAMS,usenatbib]{mnras}
\usepackage{epsfig}
\usepackage{amssymb}
\usepackage{gensymb}
\usepackage{amsfonts}
\usepackage{amsmath}
\usepackage{color}
\usepackage{lscape,longtable}
\usepackage{times}
\usepackage{tablefootnote}
\usepackage{subfig}
\usepackage{caption}

\long\def\symbolfootnote[#1]#2{\begingroup%
\def\thefootnote{\fnsymbol{footnote}}\footnote[#1]{#2}\endgroup} 
\def\aj{AJ}
\def\araa{ARA\&A}
\def\apj{ApJ}
\def\apjl{ApJ}
\def\apjs{ApJS}
\def\aap{A\&A}
\def\aapr{A\&A~Rev.}
\def\mnras{MNRAS}
\def\pasp{PASP}
\def\pasj{PASJ}
\def\nat{Nature}

\title[Three new VHS-DES quasars at \boldmath{$6.7\lesssim z \lesssim 6.9$}]
{Three new VHS-DES Quasars at \boldmath{$6.7\lesssim z \lesssim 6.9$} and 
Emission Line Properties at $z>6.5$} 
\author[S. L. Reed, M. Banerji et al.]
{\parbox{\textwidth} 
{S. L. Reed$^{1,2,3}$\thanks{E-mail: sophiereed@princeton.edu}, 
M. Banerji$^{2,3}$, G.~D. Becker$^{4}$, P.~C. Hewett$^{2}$, P. Martini$^{5,6}$, R. G. McMahon$^{2,3}$, E. Pons$^{2}$, M. Rauch$^{7}$, T.~M.~C.~Abbott$^{8}$, S.~Allam$^{9}$, J.~Annis$^{9}$, S.~Avila$^{10}$, E.~Bertin$^{11,12}$, D.~Brooks$^{13}$, E.~Buckley-Geer$^{9}$,
A.~Carnero~Rosell$^{14,15}$,  M.~Carrasco~Kind$^{16,17}$, 
J.~Carretero$^{18}$, F.~J.~Castander$^{19,20}$,
C.~E.~Cunha$^{21}$, C.~B.~D'Andrea$^{22}$, 
L.~N.~da Costa$^{15,23}$,
J.~De~Vicente$^{14}$, S.~Desai$^{24}$,
H.~T.~Diehl$^{9}$, P.~Doel$^{13}$,
A.~E.~Evrard$^{25,26}$, B.~Flaugher$^{9}$,
J.~Frieman$^{9,27}$, J.~Garc\'ia-Bellido$^{28}$, E.~Gaztanaga$^{19,28}$,
D.~Gruen$^{21,29,30}$, J.~Gschwend$^{15,23}$, G.~Gutierrez$^{9}$, D.~L.~Hollowood$^{31}$, K.~Honscheid$^{5,32}$, B.~Hoyle$^{33,34}$, D.~J.~James$^{35}$, K.~Kuehn$^{36}$,
O.~Lahav$^{13}$, M.~Lima$^{37,15}$,
M.~A.~G.~Maia$^{15,23}$, J.~L.~Marshall$^{38}$,
R.~Miquel$^{39,18}$, R.~L.~C.~Ogando$^{15,23}$,
A.~A.~Plazas$^{1,40}$, A.~Roodman$^{21,30}$,
E.~Sanchez$^{14}$, V.~Scarpine$^{9}$,
M.~Schubnell$^{26}$, S.~Serrano$^{19,20}$,
I.~Sevilla-Noarbe$^{14}$, M.~Smith$^{41}$, R.~C.~Smith$^{8}$, F.~Sobreira$^{42,15}$, E.~Suchyta$^{43}$,
M.~E.~C.~Swanson$^{17}$,
G.~Tarle$^{26}$, D.~Thomas$^{10}$,
D.~L.~Tucker$^{9}$, V.~Vikram$^{44}$
 \\ \small{Affliations at end
of paper.}  }}

\begin{document}
\begin{NoHyper}
\maketitle

\begin{abstract}
We report the results from a search for $z>6.5$ quasars using the Dark Energy Survey (DES) Year 3 dataset
combined with the VISTA Hemisphere Survey (VHS) and \textit{WISE} All-Sky Survey. Our photometric selection
method is shown to be highly efficient in identifying clean samples of high-redshift quasars leading to 
spectroscopic confirmation of three new quasars - VDESJ\ 0244$-$5008 ($z=6.724$), VDESJ\ 0020$-$3653 ($z=6.834$) and VDESJ\ 0246$-$5219 ($z=6.90$) - which were selected
as the highest priority candidates in the survey data without any need for additional follow-up observations. The new quasars span
the full range in luminosity covered by other $z>6.5$ quasar samples (J$_{\rm{AB}}=20.2$ to $21.3$; M$_{1450}=-25.6$ to $-26.6$). We have obtained spectroscopic
observations in the near infrared for VDESJ\ 0244$-$5008 and VDESJ\ 0020$-$3653 as well as our previously identified
quasar, VDESJ\ 0224$-$4711 at $z=6.50$ from Reed et al. (2017). We use the near infrared spectra to derive virial black-hole masses from the full-width-half-maximum of the MgII line. These black-hole masses are $\simeq$ 1 - 2 $\times$ 10$^9$M$_\odot$. Combining with the bolometric luminosities of these quasars of L$_{\rm{bol}}\simeq$ 1 - 3 $\times$ 10$^{47}$implies that the Eddington ratios are high - $\simeq$0.6-1.1. 
We consider the C\textrm{\textsc{IV}} emission line properties of the sample and demonstrate that our high-redshift quasars
do not have unusual C\textrm{\textsc{IV}} line properties when compared to carefully matched low-redshift samples. Our new DES+VHS $z>6.5$ quasars now add to the growing census of 
luminous, rapidly accreting supermassive black-holes seen well into the epoch of reionisation. 

\end{abstract}

\begin{keywords} dark ages, reionisation, first stars --- galaxies: active ---
galaxies: formation --- galaxies: high redshift -- quasars individual:
\end{keywords}

\section{Introduction}

The Epoch of Reionisation (EoR) represents a transformational period in the history
of the Universe when it transitioned from a predominantly neutral to a predominantly
ionised state. Luminous quasars are among the best probes of this era in the Universe's
history, and high signal-to-noise ratio (S/N), high-resolution spectra of the most luminous quasars
can be used to determine the neutral hydrogen fraction e.g. by studying the properties of 
the Ly$\alpha$ forest and the sizes of quasar proximity zones (e.g. \citealt{Fan2006, Bolton2007}). Furthermore, the
identification of such luminous quasars early in the Universe's history poses significant
challenges for theories of black-hole seed formation and growth (e.g. \citealt{Volonteri2010, Latif2013}) requiring
massive seeds as well as extended periods of Eddington-limited or super Eddington growth to explain the population 
(e.g. \citealt{Sijacki2009}). 

Around 100 luminous quasars are now known at $z\sim6-6.5$ (e.g. \citealt{Banados2016, Jiang2016, Reed2017, Wang2017}). The search for luminous quasars is now being pushed to even higher redshifts, aided by the incorporation of red-sensitive CCDs and
filters in
wide-field ``optical" surveys such as The Dark Energy Survey (DES), DECals, Pan-STARRS and HyperSuprimeCam (HSC). These improvements in area, depth and sensitivity enable quasars to be identified at $z>6.5$. 
The challenge of identifying quasars at these highest redshifts is demonstrated clearly by the fact that for 
the last seven years only a single quasar
was known above $z=7$ \citep{Mortlock2011} with the redshift record only recently broken by the $z=7.54$ quasar identified by \citet{Banados2018}.  
Identifying these most distant quasars requires the combination of wide-field optical surveys (in which the quasars
appear as drop-outs) with sensitive near infra-red surveys (in which the quasars are detected). Near infrared surveys such as 
the UKIDSS \citep{Mortlock2011, Banados2018}, VISTA Hemisphere Survey (VHS; \citealt{Venemans2015, Pons2018}) and VIKING \citep{Venemans2013} have therefore
been crucial to pushing the redshift frontiers for quasar discovery. Many of the discoveries of $z>6.5$ quasars have come within the last year with the new data from surveys such as DES, DECals and HSC in combination with near infra-red data from UKIDSS, VHS and \textit{WISE} playing a crucial part \citep{Matsuoka2018, Matsuoka2018b, Banados2018, Wang2018, Yang2018}. Identifying more quasars at these highest
redshifts is critical in order to constrain models of reionisation as well as black-hole formation and growth. 

In this paper we present our search for quasars with $z>6.5$, exploiting the wide wavelength coverage provided by combining data from 
DES, VHS and the \textit{WISE} All-Sky Survey. We also present new near infrared spectra for three of 
our four $z>6.5$ quasars. The near infra-red spectra give us access to a whole host of rest-frame UV emission lines, which trace the 
dynamics of the quasar broad-line region (BLR). We use these emission lines to derive more robust redshifts, estimate black-hole masses
as well as look for evidence for powerful disk winds affecting the BLR. 

Throughout
this paper we assume a flat $\Lambda$CDM cosmology with $\Omega_{\rm{M}}=0.3$, $\Omega_{\Lambda}=0.70$ and 
H$_0$=70.0 km s$^{-1}$ Mpc$^{-1}$. 
All magnitudes are on the AB system,
which is the native photometric system for DES. For VHS and \textit{WISE} we have used Vega to AB conversions of  $J_{\rm{AB}} = J_{\rm{Vega}} + 0.937$, $K_{\rm{S,AB}} = K_{\rm{S,Vega}} + 1.839$, $W1_{\rm{AB}} = W1_{\rm{Vega}}$ + 2.699 and  $W2_{\rm{AB}} = W2_{\rm{Vega}}$ + 3.339\footnote{\url{http://wise2.ipac.caltech.edu/docs/release/allwise/expsup/sec5\_3e.html}}.  

\section{Photometric Selection}

\subsection{Dark Energy Survey (DES)}

\label{sec:DES}

In \citet{Reed2017} (R17 hereafter) we presented the discovery of eight $z>6$ quasars identified using data from the first year of DES observations (Y1). The 10$\sigma$ depths for DES Y1 from \citet{Reed2017} are $g=24.2$, $r=23.9$, $i=23.3$, $z=22.5$ and $Y=21.2$.
In this paper we use the internal DES releases (known as Y1 and Y3) corresponding to the first three years of DES observations.
The DES Y3 release has been published as DES Data Release 1 \citep{Abbott2018} and covers $\sim$5000 deg$^2$ of the sky to 10$\sigma$ depths of $g=24.3$, $r=24.1$, $i=23.4$,  $z=22.7$ and $Y=21.4$ in a 1.95 arcsecond diameter aperture. Thus the Y3 release probes $\sim$0.2 mags deeper than the Y1 data in the $z$-band and covers almost three times the area of DES Y1.  We use the catalogues produced by the DES Collaboration throughout the paper. All DES magnitudes used in the paper are PSF magnitudes unless otherwise stated. 

\subsection{VISTA Hemisphere Survey (VHS)}

The search for quasars at the highest redshifts requires the optical data from DES
to be supplemented with near infrared photometry. In particular, observations in the
near infra-red $J$-band are important to break the degeneracy in colours between cool
stars and high redshift quasars at $6.6<z<6.8$. We therefore also make use
of photometry in the $J$ and $K_S$ bands from the VISTA Hemisphere Survey (VHS; \citealt{McMahon2013, Banerji2015}) in this work.
The VHS data used here covers $\sim$68\% of the $\sim$5000
deg$^{2}$ area of the DES Y3 data release, discussed in Section
\ref{sec:DES}. 
Thus the combined DES+VHS area within which we search for 
high-redshift quasars is $\sim$3400 deg$^2$. In this paper we make use of the VHS catalogue magnitudes measured in a two arcsec diameter aperture (\textit{apermag3}) with an appropriate aperture correction for point sources.  

\subsection{WISE All-Sky Survey}

\label{sec:wise}
Longer wavelength data at 3.4 and 4.6$\mu$m (known as the W1 and W2 bands respectively) were used from the all-sky Wide Infrared Survey Explorer dataset (\textit{WISE}; \citealt{Wright2010}). We used the un\textit{WISE} reduction of the NEOWISE-R3 images \citep{Meisner2017}. These coadd images are deeper than those in the All\textit{WISE} data release with 5$\sigma$ point source depths of W1$_{\textrm{AB}}$ = 20.2 and W2$_{\textrm{AB}}$ = 19.8. The \textit{WISE} data overlaps with the full DES+VHS area of our search and \textit{WISE} fluxes were measured by performing forced aperture photometry on the un\textit{WISE} coadds using the locations of sources from the VHS $J$-band catalogues. 

\subsection{Quasar Candidate Selection via SED-fitting}

\label{sec:sedcands}

\begin{table}
\caption{Summary of the steps in the z $>$ 6.5 quasar candidate photometric selection process.}
\begin{center}
\begin{tabular}{cccc}
\hline 
Step & Description & Number & Number \\ 
& & Removed & Remaining \\
\hline 
  & Number of objects in catalogue & & 425,880,019\\
1 & Flag criteria & & \\
  & Y$_{\mathrm{PSF}}\leq$ 21.5 and $\sigma_{\mathrm{Y}} < 0.2$ & & \\
  & z$_{\mathrm{PSF}}$ $- $Y$_{\mathrm{PSF}} >$ 0.5 & & \\
  & g$_{\mathrm{PSF}}$ and r$_{\mathrm{PSF}}$ $>$ 23.0 & & \\
  & $\sigma_{g}$ and $\sigma_{r}$ $>$ 0.1 &  &  \\
2 & Match to un\textit{WISE} forced & & 821,709 \\
  & photometry & & \\
3 & Y$-$J $<$ 1.0 & 606,347 & 215,362 \\
4 & Y $<$ 21.0 & 171,369 & 43,993 \\
5 & 6.3 $<$ z$_{\mathrm{predicted}} <$ 7.2 & 41,633 & 2,360 \\
6 & $\chi^{2}_{\mathrm{Quasar}} <$ 25.0 and $\chi^{2}_{\mathrm{BD}} >$ 2.0 & 2,081 & 279 \\
\hline
\end{tabular} \end{center}
\label{selectionTable}
\end{table}

\begin{figure}
\includegraphics[width = 1.1\columnwidth]{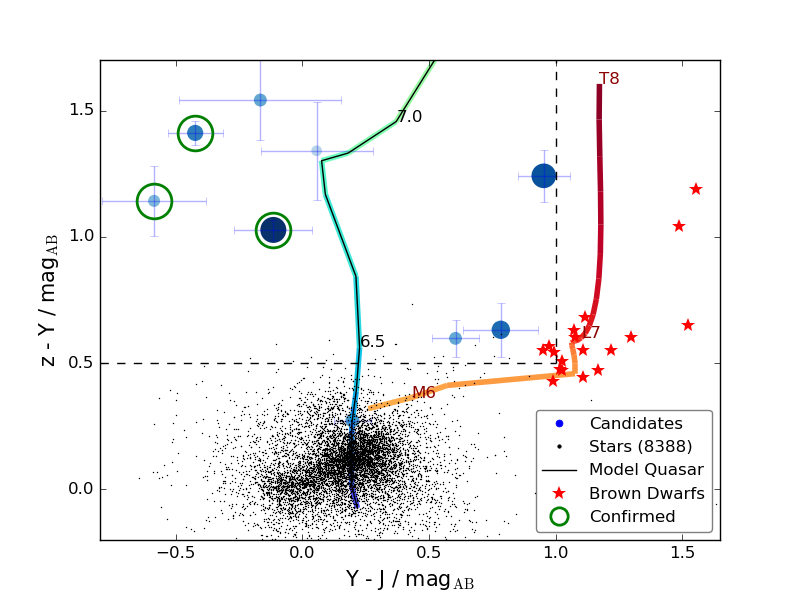}
\caption{The initial colour selection used in this paper. The yellow-red line shows the predicted path of brown dwarfs from the models used in this paper, the red stars show the colours of known brown dwarfs taken from \citet{Kirkpatrick2011} and are mostly found outside the quasar selection box which is delineated by the dashed lines. The blue-green line shows the predicted path of quasars. The blue points show the final eight quasar candidates selected after visual inspection and the size of the points is proportional to how good their fit to a quasar model is with the largest points having the best fit. The circled objects were spectroscopically followed up and confirmed to be quasars. The small black points are a sub-sample of the DES+VHS photometric sources shown for comparison to the location of the quasar candidates.}
\label{zY_YJ} \end{figure}

In order to select high-redshift quasar candidates, a series of loose flux limits and colour cuts were first applied to the combined photometric catalogues from DES + VHS + un\textit{WISE} covering $\sim$3400 deg$^2$. These cuts are summarised in Table \ref{selectionTable} in the order in which they are applied to the data. The loose colour-cuts allow us to reject sources with unphysical colours and narrow down the number of objects on which we perform full spectral energy distribution (SED) fitting. The cuts applied are broadly similar to those in R17. In that work we demonstrated that some of the high-redshift quasars recovered by our SED-fitting selection method spanned a wider range of colours compared to previous high-redshift quasar searches - e.g. VDES
J2250-5015 in R17 is too red in terms of its $(Y-J)$ colour to satisfy the selections in e.g. \citet{Banados2016} and  \citet{Venemans2015b}. To allow the inclusion of redder quasars in our sample we relaxed the $(Y-J)$ colour cut further from $<$0.8 used in R17 to $<$1.0 in this work. The initial colour selection box is shown in Figure \ref{zY_YJ} along with the predicted tracks of quasars and brown dwarfs and our final sample of photometric candidates that satisfy all the selection criteria.

\begin{figure*}
\includegraphics[width = \linewidth]{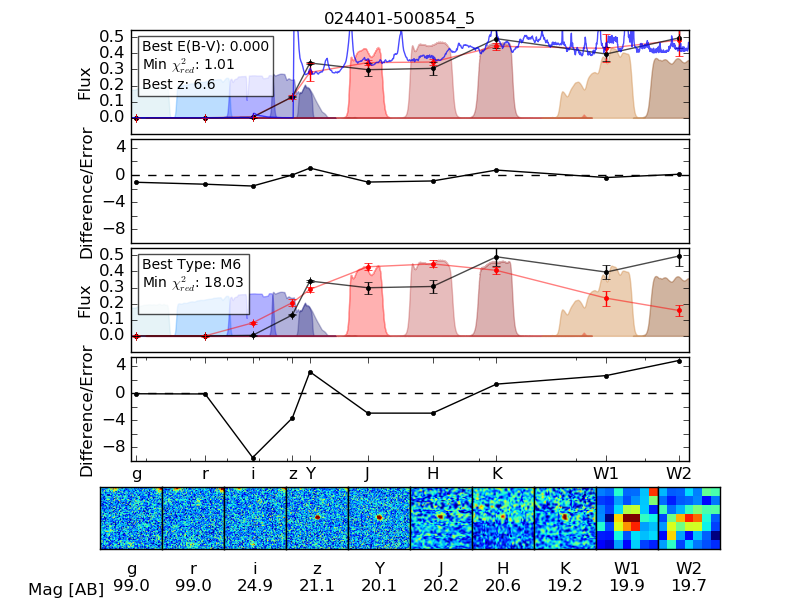}
\caption[The SED fitting method demonstrated for VDES J0244-5008 the z $>$
6.5 candidate followed up from Y3. It was found to have a spectroscopic
redshift of z $\sim$ 6.75.]{The SED fitting method illustrated for VDES J0244-5008 the first
z $>$ 6.5 candidate identified from DES data. The top two panels show
the best quasar fit and the difference (between the predicted and actual values) divided by the total error (model error and actual error combined) for this fit respectively. The next two
panels show the same for the best fitting brown dwarf model. The bottom row of
images shows 20 arcsecond cutout images across the DES, VHS and \textit{WISE}
bands. VDES J0244-5008 was confirmed as a quasar with z = 6.724.}
\label{VDESJ0244-5008_fit} \end{figure*}

The loose colour cuts and flux-limits result in a sample of 215,362 photometric candidates. These are further cut down to 43,993 by applying a brighter $Y$-band flux limit to the sample given that most optical spectrographs used for spectroscopic follow-up have relatively poor response in this wavelength range.  The SED-fitting method introduced in R17 was then applied to the photometric candidates in order to estimate the probabilities of the candidate being either a quasar or a brown-dwarf. In the case of the quasar model-fits, best-fit photometric redshifts and extinctions were also derived for each candidate. The quasar and brown-dwarf models employed were identical to those in R17 \citep{Maddox2006, Maddox2012, Skrzypek2015}. An example of the results of SED-fitting for one of our quasar candidates, VDESJ0244-5008 can be seen in Fig. \ref{VDESJ0244-5008_fit}. As the objective of this study was to identify quasars at the highest redshifts we selected candidates with a photometric redshift of $>6.3$. This study is aimed at finding quasars with $z>6.5$ but a slightly lower photometric redshift limit was used to allow for scatter in the photometric redshift estimate. Spectroscopic confirmation of our high-redshift quasar candidates makes use of optical spectrographs (see Section \ref{Q1Spec}). We therefore also imposed an upper
redshift limit of $z<7.2$, above which the Ly$\alpha$ emission line redshifts out of the range of most optical spectrographs. The candidates within this redshift range were then narrowed down based on their $\chi^{2}_{\rm{Quasar}}$ and $\chi^{2}_{{BD}}$ values, which represent the reduced $\chi^2$ values obtained from the quasar and brown-dwarf model fits respectively. Specifically, candidates with $\chi^{2}_{Quasar}>25.0$ or $\chi^{2}_{BD}<2$ were removed from the sample based on the distribution of values shown in Fig. \ref{chi2Plot}. This led to 279 high-redshift quasar candidates. All 279 candidates were visually inspected following which we identified eight candidates as the most probable high-redshift quasars. The colours of these eight candidates are shown in Fig. \ref{zY_YJ}. During the visual inspection stage, the majority of objects removed corresponded to instances of blended sources in the un\textit{WISE} coadds, which were resolved in the DES and VHS images. As the un\textit{WISE} forced photometry for these blended objects was biased artifically bright, it improved their fit to a quasar model. Other sources removed include diffraction spikes, cosmic rays and saturated objects. Of the eight remaining candidates, one (VDESJ0244-5008) had already been identified by us as a high-redshift quasar candidate using DES Y1 data and was spectroscopically followed up in January 2015 (Section \ref{Jan15spec}). Of the remaining seven candidates, five were detected in more than one VHS band and were therefore deemed higher priority. Two of the five candidates (VDESJ0020-3653 and VDESJ0246-5219) were visible during our spectroscopic observing runs (Section \ref{Q1Spec}) and were therefore followed-up. No spectroscopy has as yet been obtained for the other candidates. DES cutout images for all three high-redshift quasars with spectroscopic follow-up observations can be seen in Fig. \ref{DEScutouts} and the photometry for all three sources is summarised in Table \ref{tab:propertiesz6p75}. For completeness we also include in Table \ref{tab:propertiesz6p75} the properties of VDESJ0224-4711, which is the other $z\geq6.5$ quasar previously identified by us using DES+VHS in R17. 

\begin{figure}
\includegraphics[width = 1.1\columnwidth]{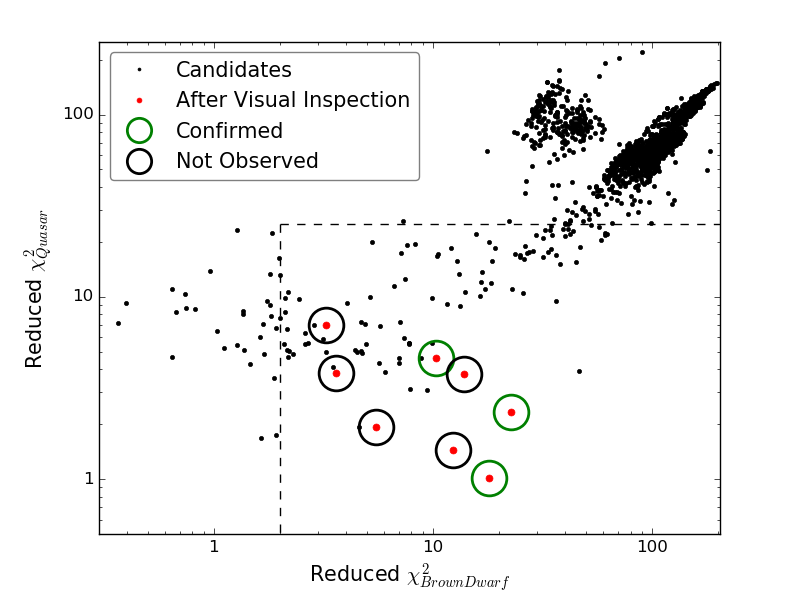}
\caption{The results of the $\chi^{2}$ SED fits for the 2360 candidates left after the predicted redshift cuts in Table \ref{selectionTable}. The dashed lines show the cuts used to narrow down the candidate list for visual inspection. These cuts remove the clearly defined locus of points that have very high $\chi^{2}$ for both the quasar and brown dwarf model fits. Visual inspection showed these to primarily be objects with contamination in the Y band, such as diffraction spikes. The final eight candidates remaining after visual inspection of all sources in the selection region are shown as the red points. The spectroscopically confirmed quasars are highlighted with green circles.}
\label{chi2Plot} \end{figure}

\begin{figure*}
\includegraphics[width = \linewidth]{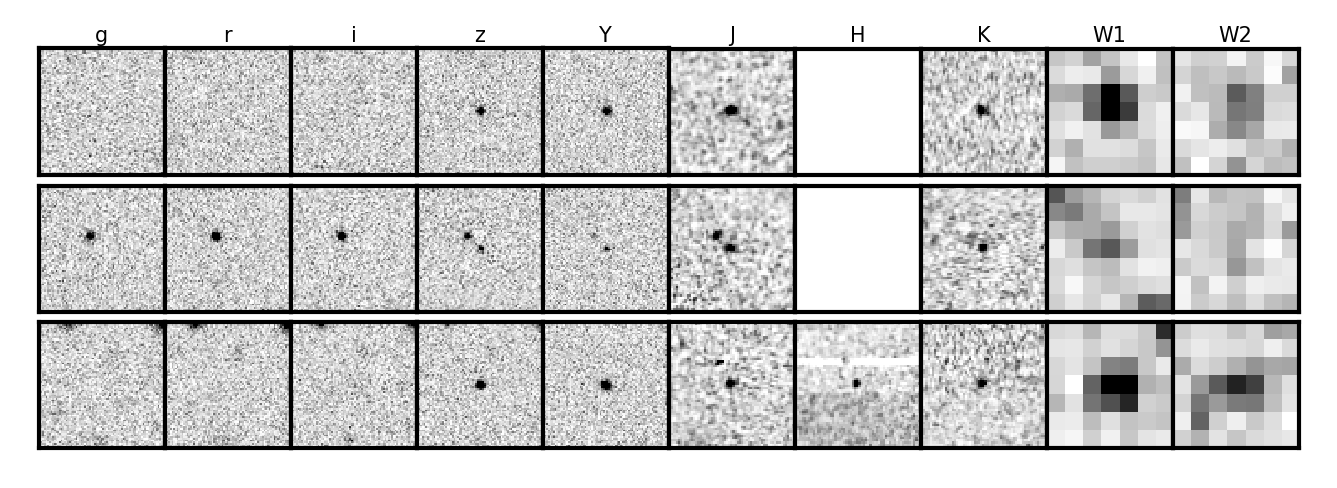}
\caption{DES optical, VHS near-IR un\textit{WISE} IR 20 arcsecond cutouts of the three new $z>6.5$ quasars identified in this paper. Top to
bottom: VDES J0020-3653, VDES J0246-5219, VDES J0244-5008. All three quasars appear as drop-outs in the DES $gri$ bands as expected. There is no $H$-band data for the top two objects. In these cutouts North is down and East is to the left.}
\label{DEScutouts}
\end{figure*}

\begin{table*} \begin{center} \caption{Positions and magnitudes of the three new z $>$ 6.5
quasars discovered in this work as well as the $z\sim6.5$ quasar VDESJ0224-4711 from R17. The g, r and i band magnitudes are given as
a 5$\sigma$ magnitude limits for a 2 arcsecond aperture.}
\label{tab:propertiesz6p75} 
\begin{tabular}{ccccc}
\hline 
& VDES0224-4711 & VDES J0244-5008 & VDES J0020-3653 & VDES J0246-5219 \\ \hline 
DES Tilename & DES0222-4706 & DES0245-4957 & DES0021-3706 & DES0246-5205 \\
RA (J2000) & 36.11057 & 41.00424 & 5.13113 & 41.73289 \\
& 02$^{h}$24$^{m}$26.54$^{s}$ & 02$^{h}$44$^{m}$01.02$^{s}$ & 00$^{h}$20$^{m}$31.47$^{s}$ & 02$^{h}$46$^{m}$55.89$^{s}$ \\
Dec. (J2000) & -47.19149 & -50.14826 & -36.89495 & -52.33054 \\
& -47$^{\circ}$11'29.4" & -50$^{\circ}$08'53.7" & -36$^{\circ}$53'41.8" & -52$^{\circ}$19'49.9"\\
g & $>$ 25.0 & $>$ 24.0 & $>$ 24.0 & $>$ 24.0 \\
r & $>$ 25.0 & $>$ 24.4 & 25.53 $\pm$ 0.60 & $>$ 24.4 \\
i & 24.0 $\pm$ 0.4 & $>$ 23.9 & 25.01 $\pm$ 0.64 & $>$ 23.9 \\
z & 20.20 $\pm$ 0.02 & 21.08 $\pm$ 0.08 & 21.39 $\pm$ 0.04 & 21.85 $\pm$ 0.11 \\
Y & 19.89 $\pm$ 0.05 & 20.15 $\pm$ 0.05 & 19.98 $\pm$ 0.03 & 20.70 $\pm$ 0.08 \\
J & 19.75 $\pm$ 0.06 & 20.21 $\pm$ 0.15 & 20.40 $\pm$ 0.10 & 21.29 $\pm$ 0.19 \\
Ks & 18.99 $\pm$ 0.06 & 19.67 $\pm$ 0.14 & 19.55 $\pm$ 0.13 & 20.35 $\pm$ 0.21 \\
W1 & 18.75 $\pm$ 0.05 & 19.91 $\pm$ 0.12 & 19.82 $\pm$ 0.14 & 20.09 $\pm$ 0.14 \\
W2 & 18.6 $\pm$ 0.1 & 19.02 $\pm$ 0.15 & 19.71 $\pm$ 0.32 & 21.89 $\pm$ 0.81 \\
\hline
\end{tabular}
\end{center}
\end{table*}

\section{Spectroscopic Observations}

\label{Q1Spec}

This section presents details of the spectroscopic observations conducted for our three $z>6.5$ 
quasar candidates identified in Section \ref{sec:sedcands}. We begin by describing
the optical spectroscopic observations used to confirm that our photometric candidates are true high redshift quasars. 
We then present near infra-red spectra that allow us to derive emission line properties and black-hole 
masses for these quasars. In addition to the three $z>6.5$ quasar candidates, we also present here new 
optical and near infra-red spectra for the $z=6.50$ quasar VDESJ0244-4711, which was first 
identified in R17. 

\begin{figure}
\includegraphics[width = \linewidth]{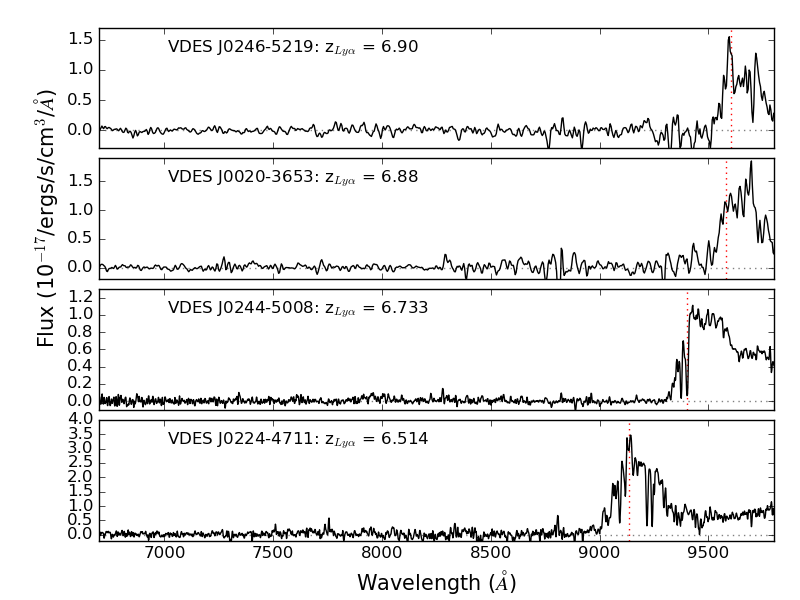}
\caption{Optical discovery spectra of the three new $z>6.5$ quasars identified in this work as well as the new Gemini GMOS spectrum of the $z\sim6.5$ quasar VDESJ0224-4711 (Section \ref{sec:VDESJ0244}) from R17.
The red dotted lines mark the derived Ly$\alpha$ redshift for these quasars and the redshifts derived from the onset of the Ly$\alpha$ emission line are indicated in the figure legend.}
\label{optSpecs} \end{figure}

\begin{table*}
\begin{center}
\caption{Observational details for the optical spectroscopy of the
three new z $>$ 6.5 quasars as well as the $z=6.50$ quasar VDESJ0224-4711 from R17.}
\label{obsSetup}
\begin{tabular}{ccccccc}
\hline
Name & Telescope & Instrument & Exposure Time & Date & Filter & Grating/ \\
& & & (Seconds) & & & Grism \\
\hline
VDES J0224-4711 & Gemini-S & GMOS & 300 $\times$ 4 = 1200 & 07/10/2016 & RG610\_G0331 & R400+\_G5325\\
VDES J0244-5008 & Clay & MagE & 600 + 1200 $\times$ 2 = 3000 & 18/01/2015 & OG-590 & VPH-Red \\
VDES J0020-3653 & NTT & EFOSC2 & 1800 + 1800 = 3600 & 25/12/2016 & OG530 & Gr\#16 \\
VDES J0246-5219 & NTT & EFOSC2 & 2400 + 2400 = 4800 & 15/11/2016 & OG530 & Gr\#16 \\
\hline
\end{tabular}
\end{center}
\end{table*}

\subsection{Optical Spectroscopy}
\label{OptSpec}

\subsubsection{Las Campanas Clay MagE}
\label{sec:VDESJ0244}

VDESJ0244-5008 was our first $z>6.5$ quasar candidate identified using DES Y1 photometry. 
In January 2015 the source was observed using the Magellan
Echellette (MagE) Spectrograph on the 6.5m Clay Telescope at Las
Campanas. Details of the observational setup can be found in Table \ref{obsSetup}. The data were reduced using a custom suite of IDL routines (e.g., \citealt{Becker2012}).  Individual frames were flat-fielded and the sky emission subtracted using an optimal b-spline fit to the sky following \citet{Kelson2003}. Relative flux calibration was performed using standard stars.  For each detector, a single one-dimensional spectrum was simultaneously extracted from all two-dimensional exposures of a given object using optimal techniques \citep{Horne1986}.  Corrections for telluric absorption were done using model transmission curves based on the ESO SKYCALC Cerro Paranal Advanced Sky Model \citep{Noll2012, Jones2013}. The MagE spectrum can be seen in the third panel of Fig. \ref{optSpecs}. The optical spectrum and the onset of the Ly$\alpha$ emission line (see R17 for details) imply a redshift of $z = 6.733 \pm 0.008$ for the quasar. Further to the method given in R17 the uncertainties on the redshifts were calculated by taking 100 realisations from the error spectrum and adding them to the spectrum before running them through the same redshift determination method. The uncertainty given is the $\sigma_{\textrm{MAD}}$ from this distribution.

\subsubsection{ESO NTT EFOSC2}

\label{nttspec}

The two candidates - VDESJ0246-5219\footnote{This quasar was recently independently discovered by \citet{Yang2018}} and VDESJ0020-3653 -  
identified from the DES Y3 data, were observed using the
European Southern Observatory's 3.6m New Technology Telescope (NTT).
Observations were taken during December 2016
and November 2017 as part of programmes 098.A-0439 and 0100.A-0346 respectively. A summary
of the observational setup is given in Table \ref{obsSetup}. The spectra were reduced
using a custom python library designed for reducing high redshift quasar
spectra and detailed in R17. Flat fielding and dark subtraction was done using calibration products taken during the afternoon preceding the observations.  Cosmic rays were removed from the image using a python implementation (cosmics.py) of the LA cosmics algorithm \citep{VanDokkum2001}. The object was then extracted from the calibrated and cleaned image using a Gaussian extraction kernel and the response function corrected for using standard star measurements. Finally the one-dimensional spectra were flux calibrated to reproduce the observed magnitudes of the object in DES and VHS. The reduced spectra can be seen in the top two panels of Fig. \ref{optSpecs}
and confirms the identity of both candidates as high-redshift quasars. Based on the onset of the Ly$\alpha$ forest we derive redshifts of $6.90\pm 0.02$ and $6.86 \pm 0.01$ for VDESJ0246-5219 and VDESJ0020-3653 respectively. 

\subsubsection{Gemini South GMOS}
\label{Jan15spec}
VDESJ0224-4711 was first identified in R17 using DES Y1 data and spectroscopically confirmed 
as a $z=6.50$ quasar via observations with the EFOSC2 spectrograph on the European Southern Observatory (ESO)'s New Technology Telescope (NTT). Here we present a new higher spectral resolution, higher S/N rest-frame UV spectrum of the same object taken with the GMOS spectrograph on Gemini-S. Details of the observational setup can be found in Table \ref{obsSetup}. The data were reduced 
using the methods outlined in R17 for the Gemini GMOS data and are broadly similar to those employed for the NTT spectral reductions (Section \ref{nttspec}). The new GMOS spectrum for VDESJ0224-4711 can be seen in the bottom panel of Fig. \ref{optSpecs}. 

The new spectrum allows for a better
estimate of the quasar ionization near zone size compared to the NTT discovery spectrum. The analysis of near zone sizes for all our high-redshift quasars will
be presented in a forthcoming paper. We also derive a new Ly$\alpha$
redshift of $z=6.514 \pm 0.005$ based on this spectrum (see R17 for details on the redshift estimation method).

\subsection{Infrared Spectroscopy}
\label{IRSpec}

\subsubsection{Gemini South Flamingos}

Near infrared spectra were obtained for VDES J0224-4711 and
VDES J0244-5008 using the Flamingos 2 (F2) spectrograph on the Gemini South
telescope. We used the long-slit spectroscopy mode with a slit
width of 4 pixels (corresponding to 0.72 arcsecs). F2 uses a 2048x2048 Hawaii-II (HgCdTe) detector with
18-micron pixels. There are two grisms used in the setup for these
observations, the JH and HK grisms. The JH grism covers 0.9 to 1.8 microns and
the HK grism covers 1.2 to 2.4 microns.  These can then be combined to give
coverage from 0.9 to 2.4 microns in the observed frame.

\begin{figure*}
\includegraphics[width = \linewidth]{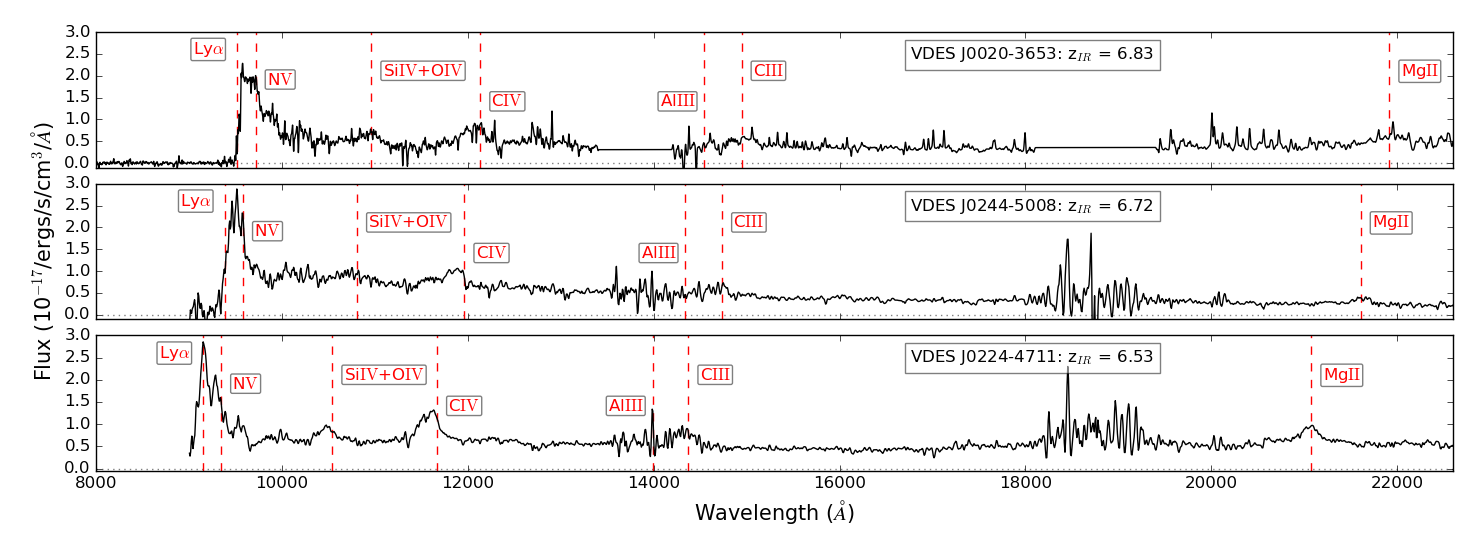}
\caption{Near infrared spectra of three of our four z$>$6.5 quasars. The
red lines show the positions of spectral lines in the observed frame assuming the Mg\textrm{\textsc{II} derived systemic redshifts from Table \ref{linefit}. The areas of very noisy (or blanked) flux, around 13800\AA\@ and 19000\AA have very little flux due to the limited transmission at those wavelengths.}
\label{IRSpecs}}
\end{figure*}

VDES J0224-4711 was observed on November 2016 and VDES J0244-5008 was
observed on January 2017. Both objects were observed for 8$\times$200 second exposures 
in both JH and HK. The observations were taken in pairs using an ABBA
nodding pattern. Each pair was then
reduced together: one observation was subtracted from the other to remove
artifacts and as a first pass at sky subtraction. The subtracted image was then
flat fielded to leave a relatively clean image with a positive and negative trace visible. 
Each trace was extracted separately and later combined. Wavelength calibration was performed using Argon lines from arc lamp spectra taken
during the night of the observations. A 5$^{\mathrm{th}}$ order Chebyshev polynomial fit was used to determine the wavelength solution. After the 
wavelength calibration the median of the eight individual exposures was used as the 
final output spectrum. 

The system response was calculated using
observations of the A0V type standard star HIP6364, taken just preceding
the observations of the target.  The standard star observations were reduced in the
same way as the target data with the only difference being that two pairs of
observations were taken rather than four. The output spectrum of the star was then divided by an A0V spectral template\footnote{http://axe.stsci.edu/html/templates.html} in order to give the telluric and instrument response correction. Both the near infrared spectra can be seen in the bottom two panels of Fig. \ref{IRSpecs}. 

\subsubsection{VLT XShooter}
VDES J0020-3653 was observed with the XShooter spectrograph on the ESO Very
Large Telescope sited at Paranal Observatory in October 2017.
The observations were reduced using a custom set of IDL routines \citep{Lopez2016}. The data reduction steps are broadly similar to those described in Section \ref{sec:VDESJ0244}
We did not nod-subtract the X-Shooter NIR frames.  
Instead, a high S/N composite dark frame was subtracted from each exposure to mitigate the effects of dark current, hot pixels, and other artifacts prior to fitting the sky. 
The sky model used is again described in Section \ref{sec:VDESJ0244}. 
The final XShooter near infrared spectrum can be seen in the top panel of Fig. \ref{IRSpecs}. 

\section{Emission Line Properties}

We now consider the emission line properties derived from our near infrared spectra in order to constrain the systemic redshifts, 
black-hole masses and broad-line region outflow velocities of our
high-redshift quasars. The emission lines detected in the near infra-red spectra are generally of modest
S/N at the native spectral resolution of these observations. For the purposes of measuring
broad emission line properties, high spectral resolution is not a pre-requisite. We therefore create 
inverse-variance weighted binned spectra of our quasars before spectral fitting.

Line properties are derived from the binned spectra by fitting Gaussian profiles to the broad emission lines after subtraction 
of a pseudo-continuum, which is made up of a power-law component to model emission from the quasar accretion disk and 
an FeII template from \citet{Vestergaard2001}. Given the modest S/N in the continuum, an FeII template results in 
an improved fit to the MgII line only for the lowest redshift quasar, VDESJ0224-4711. In order to model the emission line, 
we begin by fitting a single Gaussian to the line profile and add additional Gaussians only if they are strongly evidenced
by the data and result in an improvement in the reduced $\chi^2$ of the fit by $>$10 per cent. 

Uncertainties on the line properties are calculated by generating 100 realisations of the spectra with the flux at each
wavelength drawn from a normal distribution with a mean value taken from the best-fit Gaussian model and a standard 
deviation given by the noise spectrum. The line-fitting is then run on each of these 100 synthetic spectra and the 
standard deviation of the resulting line-fit parameters are quoted as our formal uncertainties.   

\begin{figure*}
\begin{tabular}{cc}
\large{\textbf{C\textrm{\textsc{IV}}}} & \large{\textbf{MgII}} \\
\includegraphics[scale=0.4]{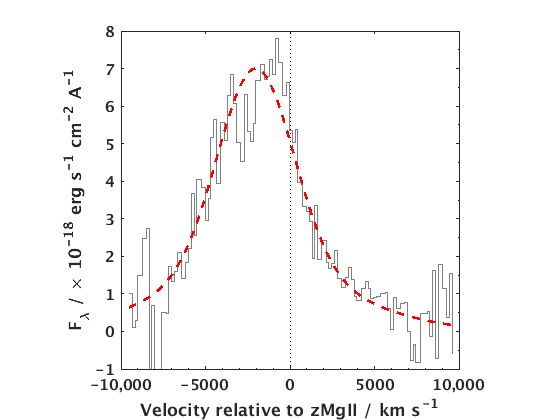} & \includegraphics[scale=0.4]{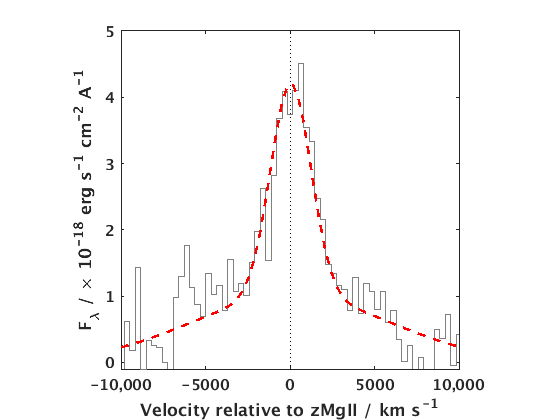} \\
\includegraphics[scale=0.4]{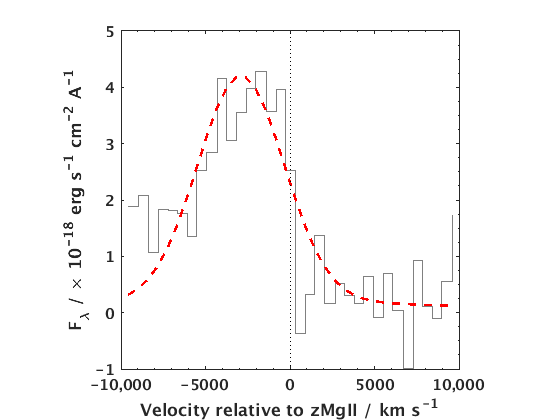} & \includegraphics[scale=0.4]{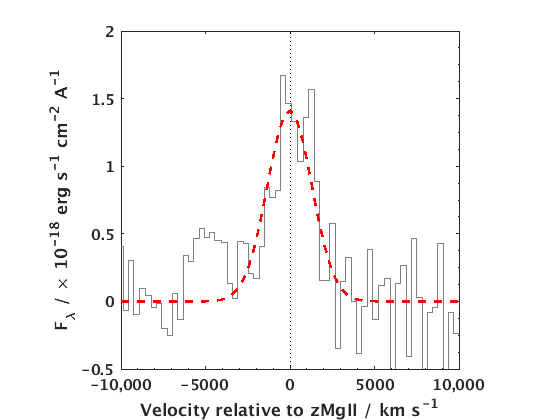} \\
\includegraphics[scale=0.4]{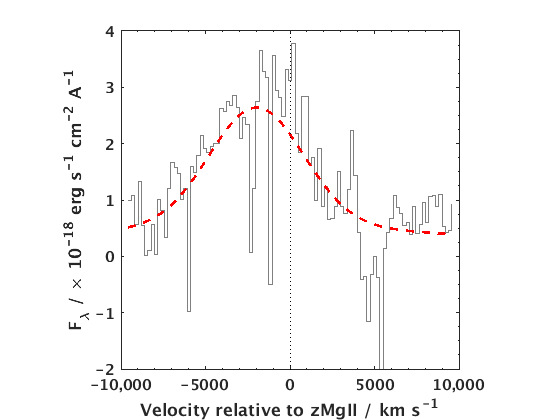} & \includegraphics[scale=0.4]{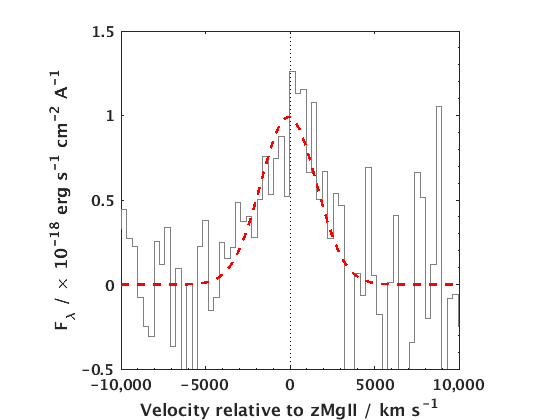} \\
\end{tabular}
\caption{Gaussian fits to the C\textrm{\textsc{IV}} (left) and MgII (right) emission line profiles for VDESJ0224-4711 (top), VDESJ0244-5008 (middle) and VDESJ0020-3653 (bottom). Zero velocity is defined as the MgII peak redshift and shown by the dotted vertical lines.}
\label{linefit}
\end{figure*}

\subsection{Systemic Redshifts}

\label{MgIIz}

Robust measures of quasar ionization near zone sizes rely on an accurate estimate of the quasar systemic redshift. 
It is well known that redshift estimates based on the Ly$\alpha$ emission line can have large systematic offsets as this resonant line is affected by absorption and the kinematics therefore strongly depend on the geometry and distribution of the obscuring material, which affect the scattering of Ly$\alpha$ photons. Redshift estimates based on low-ionization rest-frame optical emission lines such as
MgII, 2798\AA\@ on the other hand are generally considered more robust \citep{Hewett2010, Shen2016}. Here we make use of our new near infrared spectra to derive systemic redshifts based on MgII and compare them to the Ly$\alpha$ redshifts presented in 
Section \ref{OptSpec}. 

The Gaussian fits to the continuum subtracted MgII line profiles for all 3 quasars can be seen in Fig. \ref{linefit}. While a single Gaussian provides a
reasonable fit to VDESJ0244-5008 and VDESJ0020-3653, in the case of VDESJ0224-5711 we find that 2 Gaussians
constrained to have the same centroid are necessary in order to adequately fit the broad wings seen in the emission 
line profile of this object. There is no evidence for a velocity offset between the two Gaussian components in this quasar and we therefore do not allow the centroid of the second Gaussian to be an additional free parameter in the fit. 
 
 From these Gaussian fits we infer MgII redshifts of 6.526$\pm$0.003, 6.724$\pm$0.002 and 6.834$\pm$0.004 for VDESJ0224-4711,
 VDESJ0244-5008 and VDESJ0020-3653 respectively. For VDESJ0244-5008, the redshift estimate is consistent with that based
 on the onset of Ly$\alpha$ but for the other two quasars, Ly$\alpha$ is redshifted by $\delta z\sim 0.01-0.03$ relative to MgII. 

\subsection{Bolometric Luminosities, Black Hole Masses \& Eddington Ratios}

We calculated bolometric luminosities for our quasars from the rest-frame 3000\AA\, luminosities assuming a bolometric correction of 5.15 \citep{DeRosa2011}. The rest-frame 3000\AA\, luminosities have been calculated by fitting our quasar SED models (Section \ref{sec:sedcands}) to the available photometry for each quasar and fixing the redshift of the model to the spectroscopic redshift of the quasar estimated from the Mg\textrm{\textsc{II}} line. Both values are quoted in Table \ref{BHMasses} and the errors are estimated by propagating the errors on the measured photometry.

Black hole masses were calculated from the full-width-half-maximum (FWHM) of the Mg\textrm{\textsc{II}} line and using the calibration in \citet{Vestergaard2009}:

\begin{equation}
\frac{M_{\rm{BH}}}{M_{\odot}}=10^{6.86} \left(\frac{\rm{FWHM}_{\rm{MgII}}}{1000 \rm{km s}^{-1}}\right)^2 \left(\frac{\rm{L}_{3000}}{10^{44} \rm{erg s}^{-1}}\right)^{0.5}
\label{eq:MBH}
\end{equation}

\noindent We derived the FWHM of the Mg\textrm{\textsc{II}} from the best-fit Gaussian model and subtracted the instrumental
resolution in quadrature from this value. Uncertainties were calculated using the 100 realisations of the line profile with noise added. The FWHM of the Mg\textrm{\textsc{II}} line together with the derived black-hole masses are given in Table \ref{BHMasses}. All three quasars have black-hole masses of $\simeq$1-2$\times$10$^9$M$_\odot$. The typical systematic uncertainties on these black-hole mass estimates are $\sim$0.3 dex \citep{Shen2018}. Combining with their bolometric luminosities of $\simeq 1-3 \times $10$^{47}$ erg s$^{-1}$ we infer Eddington ratios of close to or just above unity for all three quasars consistent with them being seen during a high-accretion growth phase. In Fig. \ref{MBHEdd} we compare the bolometric luminosities and black-hole masses to other $z>6$ quasars from the literature where such observations have been made \citep{DeRosa2011, DeRosa2014, Mazzucchelli2017, Banados2018}. Our three new quasars have bolometric luminosities, black-hole masses and Eddington ratios that are broadly consistent with other high-redshift quasars.

\begin{table*}
\begin{center}
\caption{Emission line properties of our three $z>6.5$ quasars with near infra-red spectroscopy. } \label{BHMasses}
\begin{tabular}{cccc}
\hline 
 & VDES J0020-3653 & VDES J0244-5008 & VDES J0224-4711 \\
\hline
Ly$\alpha$ Redshift & 6.86 $\pm$ 0.01 & 6.733 $\pm$ 0.008 & 6.514 $\pm$ 0.005\\
Mg\textsc{II} Redshift & 6.834 $\pm$ 0.0004  & 6.724 $\pm$ 0.0008 & 6.526 $\pm$ 0.0003 \\[5pt]
FWHM$_{\rm{MgII}}$ / kms$^{-1}$ & 3800 $\pm$ 360 & 3100 $\pm$ 530 & 3500 $\pm$ 310 kms \\
$\lambda$L$_{\lambda}$(3000) / ergs$^{-1}$ & (2.62$\pm$0.05)$\times10^{46}$ & (2.79$\pm$0.05)$\times10^{46}$ & (6.08$\pm$0.09)$\times10^{46}$ \\
M$_{\rm{BH}}$ / M$_\odot$ & (1.67$\pm$0.32)$\times10^9$ & (1.15$\pm$0.39)$\times10^9$ & (2.12$\pm$0.42)$\times10^9$ \\
L$_{\rm{bol}}$ / ergs$^{-1}$ & (1.35$\pm$0.03)$\times10^{47}$ & (1.44$\pm$0.02)$\times10^{47}$ & (3.13$\pm$0.04)$\times10^{47}$ \\
L$_{\rm{bol}}$/L$_{\rm{Edd}}$ & 0.62$\pm$0.12 & 0.96$\pm$0.33 & 1.13$\pm$0.22 \\
C\textsc{IV} Blueshift & 1700 $\pm$ 100 km$^{-1}$ & 3200 $\pm$ 310 kms$^{-1}$ & 2000 $\pm$ 160 kms$^{-1}$ \\
C\textsc{IV} EW (Restframe) & 55 $\pm$ 1  \AA  & 24 $\pm$ 2 \AA & 44 $\pm$ 2 \AA \\
\hline
\end{tabular}
\end{center}
\end{table*}
 
\begin{figure}
\includegraphics[width = \linewidth]{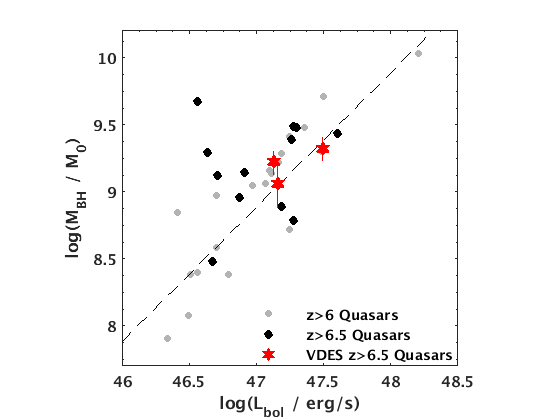}
\caption{Bolometric luminosity versus black-hole mass for the three $z>6.5$ quasars in this paper (red symbols). These are compared to $z>6$ and $z>6.5$ quasars from the literature \citep{DeRosa2011, DeRosa2014, Mazzucchelli2017}. The dashed line denotes an Eddington ratio of 1.}
\label{MBHEdd} 
\end{figure}

\subsection{C\textsc{IV} Blueshifts}

The C\textrm{\textsc{IV}} 1550\AA\@ emission line in luminous quasars has long been known 
to display systematic velocity offsets of several thousand km/s blue-ward of systemic \citep{Richards2002, Baskin2005}
which are widely thought
to be indicative of outflowing gas in the quasar broad-line region \citep{Konigl1994, Murray1995}. Attention 
has been drawn to the large C\textrm{\textsc{IV}} blueshifts seen in the spectra of the 
highest redshift quasars \citep{DeRosa2011, Mazzucchelli2017}, which could indicate
that strong disk winds are particularly prevalent in these systems. Changes in the C\textrm{\textsc{IV}} 
emission line properties of quasars - i.e. blueshift and equivalent width - are themselves correlated with 
the velocity widths and strengths of other optical and UV emission lines as well as the bolometric luminosity of the quasar \citep{Richards2011}. Recently \citet{Coatman2016}
demonstrated that $z\sim2$ quasars with high C\textrm{\textsc{IV}} blueshifts also exhibit high Eddington ratios. It is therefore interesting to explore
the C\textrm{\textsc{IV}} emission line properties of our quasars in the context of these previous observations. 

We fit the C\textrm{\textsc{IV}} emission lines in our three high-redshift quasars after subtracting a power-law continuum from the spectrum. FeII emission 
is less strong in this region compared to the MgII region of the spectrum and given the 
typical S/N of these spectra, we did not consider it necessary to include an FeII component
in the continuum. Each component of the C\textrm{\textsc{IV}} doublet is modeled as the sum of two Gaussians
with a fixed velocity separation between the doublet components of 390 km/s.  The use of two Gaussians to model each component of the doublet allows us to adequately reproduce the asymmetric C\textrm{\textsc{IV}} line profiles, given the typical S/N of our spectra. 
The C\textrm{\textsc{IV}} blueshifts are derived from the velocity centroid of the C\textrm{\textsc{IV}} emission line relative to the
MgII derived systemic redshifts presented in Section \ref{MgIIz}. These
blueshifts range from 1700 km/s in VDESJ0020-3653 to 3200 km/s in VDESJ0244-5008 (Table 
\ref{BHMasses}). The C\textrm{\textsc{IV}} equivalent widths are also summarised in Table 
\ref{BHMasses}. We also calculated C\textrm{\textsc{IV}} line properties in an analogous way for the $z=6.82$ quasar
VHSJ0411$-$0907 recently discovered by \citet{Pons2018} deriving a blueshift of 830$\pm$20 km/s and a rest-frame equivalent
width of 32$\pm$1\AA\, for this quasar. 

In Fig. \ref{CIV_BS_EW} we compare the C\textrm{\textsc{IV}} blueshifts in our high-redshift sample with a sample
of low-redshift quasars from SDSS \citep{Shen2011}, where the low-redshift C\textrm{\textsc{IV}} blueshifts have been calculated in an analogous way to the $z>6.5$ quasars - see Section 3.2 of \citet{Coatman2016}. Specifically, the C\textrm{\textsc{IV}} emission line properties for the SDSS quasars were derived using systemic redshift
estimates using an Independent Component Analysis (ICA) technique from Allen \& Hewett (in preparation), which do not themselves include the C\textrm{\textsc{IV}} line in the systemic redshift estimate - see \citet{Coatman2016, Coatman2017} for a detailed discussion on
this issue. The ICA redshifts are completely consistent with the MgII redshifts for SDSS quasars, as well as for our high-redshift sample. However, using the ICA redshifts does allow us to expand the SDSS comparison sample to $z>2.2$, where MgII is no longer present in the SDSS spectrum. Thus our SDSS low-redshift comparison sample is much larger than those used in previous works e.g. \citet{Mazzucchelli2017}. We also note that unlike some other works in the literature we make use of the CIV velocity centroid rather than the peak velocity for all blueshift measurements. As the CIV emission line can have significant flux in the wings of the line, the centroid measure generally results in larger blueshifts compared to the peak. 

As a result of these updates, a much larger fraction of the low-z SDSS quasars now display significant
C\textrm{\textsc{IV}} blueshifts that are comparable to the high-redshift population. Our $z>6.5$ quasars (as well as those studied e.g.
by \citealt{Mazzucchelli2017}) are also among the highest luminosity, highest Eddington ratio quasars compared to the SDSS population and 
therefore expected to have large C\textrm{\textsc{IV}} blueshifts compared to the average SDSS quasar. If we select SDSS quasars with 
log$_{10}$(L$_{\rm{bol}} / \rm{ergs}^{-1})>47.0$ only and compare them to the blueshifts and equivalent widths of our four $z>6.5$ quasars we
find that a two-dimensional Kolmogorov-Smirnov test is consistent with the low and high-redshift quasar populations being drawn from the 
same continuous distribution. Very recently \citet{Shen2018} reached the same conclusion by comparing the C\textrm{\textsc{IV}} emission line properties of a large sample of $5.7\lesssim z \lesssim 6.4$ quasars to lower redshift quasars from SDSS. We have deliberately not included the \citet{Mazzucchelli2017} quasars in our test as we cannot confirm that 
the same line-fitting prescriptions have been used to calculate C\textrm{\textsc{IV}} emission line properties for these quasars as we have 
done here both for the high-redshift and low-redshift SDSS quasars. At face-value however there are three quasars in \citet{Mazzucchelli2017}
with very large blueshifts of $>$4000 km/s, which would seem inconsistent with being drawn from the same distribution as the low redshift
SDSS quasars. 

\begin{figure}
\includegraphics[width = \linewidth]{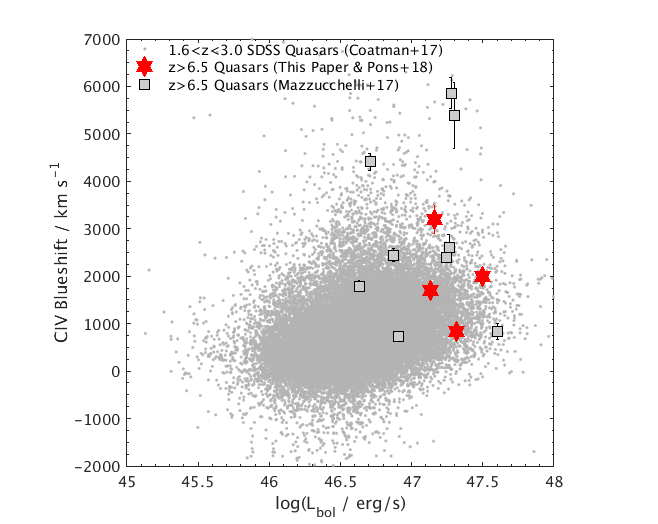}
\caption{C\textrm{\textsc{IV}} blueshift versus quasar luminosity for the three $z>6.5$ quasars in this paper and the $z=6.82$ quasar from \citet{Pons2018} (red symbols). These are compared to low-redshift quasars from SDSS as well as the $z>6.5$ quasars studied in \citet{Mazzucchelli2017}.} 
\label{CIV_BS_EW} 
\end{figure}


 
\section{Discussion and Conclusions}

We have described the discovery of three new quasars at $6.7 \lesssim z \lesssim 6.9$ identified using imaging data from the Dark Energy Survey Year 3 data release,
VISTA Hemisphere Survey and \textit{WISE}. These discoveries show that our SED-fitting method to identify $z\sim6$ quasars from wide-field imaging surveys (R17) can easily be adapted to produce clean, low-contamination samples at even higher redshifts of $z>6.5$. The three new quasars have J$_{\rm{AB}}=20.2$ to $21.3$ (M$_{1450}=-25.6$ to $-26.6$) and 
span the full luminosity range covered by other quasar samples at these high redshifts. They are at redshifts of 6.724, 6.834 and 6.90. 


We obtained near infra-red spectra for three of the four $z\ge6.5$ quasars identified by us using DES+VHS, to
constrain their black-hole masses, Eddington ratios and C\textrm{\textsc{IV}} blueshifts. The systemic redshifts derived from the MgII emission lines in these near infrared spectra are 6.526, 6.724 and 6.834. In two out of the three quasars the Ly$\alpha$ emission line is redshifted by $\delta z \sim$ 0.01-0.03 relative to MgII. Our new quasars have black-hole masses of $\simeq$ 1 - 2 $\times 10^9$M$_\odot$ 
and are accreting close to or above the Eddington limit, with derived Eddington ratios of $\sim$0.6-1.1 in the sample. This is broadly consistent with what is found for other quasars at these highest redshifts \citep{Mazzucchelli2017, Banados2018}. 

Several of our $z>6.5$ quasars exhibit large C\textrm{\textsc{IV}} blueshifts of several thousand km/s. We have demonstrated however that
if we compare the sample to lower redshift SDSS quasars of similar luminosity and where the C\textrm{\textsc{IV}} blueshift is measured in an analogous way to the high-redshift population, the distribution of C\textrm{\textsc{IV}} blueshifts and equivalent widths in our $z>6.5$ sample is completely 
consistent with the low-redshift population. Therefore it appears that high-mass, high accretion rate quasars have very similar broad-line region outflow properties regardless of the epoch at which they are observed.  

Overall our new quasars now add to the growing census of high-luminosity, highly-accreting supermassive black holes seen well into the Epoch of Reionisation. Based on extrapolations of the $z\sim6$ luminosity functions from \citet{Willott2010} and \citet{Jiang2016} we expect to find $\sim$15-20 quasars at $6.5 < z < 7.2$ down to a DES Y-band flux limit of $<$21.0 and over the full DES survey area of 5000 sq-deg. Thus the four new quasars identified in this paper are expected to form only a small subset of the $z>6.5$ quasars that will be identified using the final DES+VHS data releases.

\section{Acknowledgments}
The authors would like to thank Y. Shen for helpful comments and discussions.
SLR, RGM, PCH and EP acknowledge the support of the UK Science and Technology Facilities Council (STFC). MB acknowledges funding from the STFC via an Ernest Rutherford Fellowship as well as funding from the Royal Society via a University Research Fellowship. Support by ERC Advanced Grant 320596 ``The Emergence of Structure during the Epoch of reionization" is gratefully acknowledged by RGM. This material is based upon work supported by the National Science Foundation under Grant No. 1615553 to PM. 

The analysis presented here is based on observations obtained
as part of the VISTA Hemisphere Survey, ESO Progamme, 179.A2010
(PI: McMahon). The analysis presented here is based on observations
obtained as part of ESO Progammes, 098.A-0439 and 0100.A00346 (PI: McMahon). Based on observations obtained at the Gemini Observatory (Program GS-2016B-FT-8), which is operated by the Association of Universities for Research in Astronomy, Inc., under a cooperative agreement with the NSF on behalf of the Gemini partnership: the National Science Foundation (United States), the National Research Council (Canada), CONICYT (Chile), Ministerio de Ciencia, Tecnolog\'{i}a e Innovaci\'{o}n Productiva (Argentina), and Minist\'{e}rio da Ci\^{e}ncia, Tecnologia e Inova\c{c}\~{a}o (Brazil).

Funding for the DES Projects has been provided by the US Department of Energy,       
the US National Science Foundation, the Ministry of Science and Education of         
Spain, the Science and Technology Facilities Council of UK, the Higher               
Education Funding Council for England, the National Center for Supercomputing        
Applications at the University of Illinois at Urbana-Champaign, the Kavli            
Institute of Cosmological Physics at the University of Chicago, Financiadora         
de Estudos e Projetos, Funda{\c c}{\~a}o Carlos Chagas Filho de Amparo {\'a}         
Pesquisa do Estado do Rio de Janeiro, Conselho Nacional de Desenvolvimento           
Cient{\'i}fico e Tecnologico and the Minist{\'e}rio da Ci{\^e}ncia                   
e Tecnologia, the Deutsche Forschungsgemeinschaft and ˆ the Collaborating            
Institutions in the Dark Energy Survey.                                              
                                                                                     
The Collaborating Institutions are Argonne National Laboratories, the                
University of California at Santa Cruz, the University of Cambridge, Centro de       
Investigaciones Energeticas, Medioambientales y Tecnologicas-Madrid, the             
University of Chicago, University                                                    
College London, the DES-Brazil Consortium, the Eidgenossische Technische             
Hochschule (ETH) Zurich, Fermi National Accelerator Laboratory, the University       
of Edinburgh, the University of Illinois at Urbana-Champaign, the Institut de        
Ciencies de l’Espai (IEEC/CSIC), the Institut de Fisica d’Altes Energies, the        
Lawrence Berkeley National Laboratory, the Ludwig-Maximilians Universit{\"a}t        
and the associated Excellence Cluster Universe, the University of Michigan, the      
National Optical Astronomy Observatory, the University of Nottingham, The Ohio       
State University, the University of Pennsylvania, the University of Portsmouth,      
SLAC National Laboratory, Stanford University, the University of Sussex, and         
Texas A\&M University.

This analysis makes use of the cosmics.py algorithum based
on the L.A. Cosmic algorithm detailed in \citet{VanDokkum2001}.

\bibliographystyle{mnras}
\bibliography{Refs}

\section*{Affiliations}
{\small
$^{1}$Department of Astrophysical Sciences, 4 Ivy Lane, Princeton University, Princeton, NJ 08544\\
$^{2}$Institute of Astronomy, University of Cambridge, Madingley Road, Cambridge CB3 0HA, UK\\
$^{3}$Kavli Institute for Cosmology, University of Cambridge, Madingley Road, Cambridge CB3 0HA, UK\\
$^{4}$Department of Physics and Astronomy, University of California, 900 University Avenue, Riverside, CA 92521, USA\\
$^{5}$Center for Cosmology and Astro-Particle Physics, The Ohio State University, Columbus, OH 43210, USA\\
$^{6}$Department of Astronomy, The Ohio State University, Columbus, OH 43210, USA\\
$^{7}$Observatories of the Carnegie Institution for Science, 813 Santa Barbara
Street, Pasadena, CA 91101, USA\\
$^{8}$Cerro Tololo Inter-American Observatory, National Optical Astronomy Observatory, Casilla 603, La Serena, Chile\\
$^{9}$Fermi National Accelerator Laboratory, P. O. Box 500, Batavia, IL 60510, USA\\
$^{10}$Institute of Cosmology and Gravitation, University of Portsmouth, Portsmouth, PO1 3FX, UK\\
$^{11}$CNRS, UMR 7095, Institut d'Astrophysique de Paris, F-75014, Paris, France\\
$^{12}$Sorbonne Universit\'es, UPMC Univ Paris 06, UMR 7095, Institut d'Astrophysique de Paris, F-75014, Paris, France\\
$^{13}$Department of Physics \& Astronomy, University College London, Gower Street, London, WC1E 6BT, UK\\
$^{14}$Centro de Investigaciones Energ\'eticas, Medioambientales y Tecnol\'ogicas (CIEMAT), Madrid, Spain\\
$^{15}$Laborat\'orio Interinstitucional de e-Astronomia - LIneA, Rua Gal. Jos\'e Cristino 77, Rio de Janeiro, RJ - 20921-400, Brazil\\
$^{16}$Department of Astronomy, University of Illinois at Urbana-Champaign, 1002 W. Green Street, Urbana, IL 61801, USA\\
$^{17}$National Center for Supercomputing Applications, 1205 West Clark St., Urbana, IL 61801, USA\\
$^{18}$Institut de F\'{\i}sica d'Altes Energies (IFAE), The Barcelona Institute of Science and Technology, Campus UAB, 08193 Bellaterra (Barcelona) Spain\\
$^{19}$Institut d'Estudis Espacials de Catalunya (IEEC), 08034 Barcelona, Spain\\
$^{20}$Institute of Space Sciences (ICE, CSIC),  Campus UAB, Carrer de Can Magrans, s/n,  08193 Barcelona, Spain\\
$^{21}$Kavli Institute for Particle Astrophysics \& Cosmology, P. O. Box 2450, Stanford University, Stanford, CA 94305, USA\\
$^{22}$Department of Physics and Astronomy, University of Pennsylvania, Philadelphia, PA 19104, USA\\
$^{23}$Observat\'orio Nacional, Rua Gal. Jos\'e Cristino 77, Rio de Janeiro, RJ - 20921-400, Brazil\\
$^{24}$Department of Physics, IIT Hyderabad, Kandi, Telangana 502285, India\\
$^{25}$Department of Astronomy, University of Michigan, Ann Arbor, MI 48109, USA\\
$^{26}$Department of Physics, University of Michigan, Ann Arbor, MI 48109, USA\\
$^{27}$Kavli Institute for Cosmological Physics, University of Chicago, Chicago, IL 60637, USA\\
$^{28}$Instituto de Fisica Teorica UAM/CSIC, Universidad Autonoma de Madrid, 28049 Madrid, Spain\\
$^{29}$Department of Physics, Stanford University, 382 Via Pueblo Mall, Stanford, CA 94305, USA\\
$^{30}$SLAC National Accelerator Laboratory, Menlo Park, CA 94025, USA\\
$^{31}$Santa Cruz Institute for Particle Physics, Santa Cruz, CA 95064, USA\\
$^{32}$Department of Physics, The Ohio State University, Columbus, OH 43210, USA\\
$^{33}$Max Planck Institute for Extraterrestrial Physics, Giessenbachstrasse, 85748 Garching, Germany\\
$^{34}$Universit\"ats-Sternwarte, Fakult\"at f\"ur Physik, Ludwig-Maximilians Universit\"at M\"unchen, Scheinerstr. 1, 81679 M\"unchen, Germany\\
$^{35}$Harvard-Smithsonian Center for Astrophysics, Cambridge, MA 02138, USA\\
$^{36}$Australian Astronomical Optics, Macquarie University, North Ryde, NSW 2113, Australia\\
$^{37}$Departamento de F\'isica Matem\'atica, Instituto de F\'isica, Universidade de S\~ao Paulo, CP 66318, S\~ao Paulo, SP, 05314-970, Brazil\\
$^{38}$George P. and Cynthia Woods Mitchell Institute for Fundamental Physics and Astronomy, and Department of Physics and Astronomy, Texas A\&M University, College Station, TX 77843,  USA\\
$^{39}$Instituci\'o Catalana de Recerca i Estudis Avan\c{c}ats, E-08010 Barcelona, Spain\\
$^{40}$Jet Propulsion Laboratory, California Institute of Technology, 4800 Oak Grove Dr., Pasadena, CA 91109, USA\\
$^{41}$School of Physics and Astronomy, University of Southampton,  Southampton, SO17 1BJ, UK\\
$^{42}$Instituto de F\'isica Gleb Wataghin, Universidade Estadual de Campinas, 13083-859, Campinas, SP, Brazil\\
$^{43}$Computer Science and Mathematics Division, Oak Ridge National Laboratory, Oak Ridge, TN 37831\\
$^{44}$Argonne National Laboratory, 9700 South Cass Avenue, Lemont, IL 60439, USA\\

}

\end{NoHyper}
\end{document}